\documentclass{cjpsuppl}% for LaTeX 2e
\usepackage{epsfig}
\begin{document}
\title{Lepton Number Violation \\ in Decays of Supersymmetric Particles}
\authori{W.~Porod}
\addressi{AHEP Group, 
IFIC -  CSIC, Universitat de Val{\`e}ncia, E-46071 Valencia, Espa\~na}
\authorii{}    \addressii{}
\authoriii{}   \addressiii{}
\authoriv{}    \addressiv{}
\authorv{}     \addressv{}
\authorvi{}    \addressvi{}
\headtitle{Lepton Number Violation  in Decays of Supersymmetric Particles}
\headauthor{W.~Porod}
\lastevenhead{W.~Porod:
              Lepton Number Violation  in Decays of Supersymmetric Particles}
%%%%%%%%%%%%%% Pro editory supplementu: %%%%%%%%%%%%%%%
\refnum{}%slouzi editorum pro evidenci; nakonec {}
\daterec{} % 20 October 2002;\\final version 31 December 2003}
\suppl{A}  \year{2004} \setcounter{page}{1}
%\firstpage{1}
%\lastpage{000}
%\makefirsttitle
%%%%%%%%%%%%%%%%%%%%%%%%%%%%%%%%%%%%%%%%%%%%%%
\maketitle

\begin{abstract}
We discuss lepton flavour violating signals at the LHC in the
framework of supersymmetric theories. We consider R-parity
conserving as well as R-parity violating scenarios. In the  case of R-parity
conservation
we show that in decays of supersymmetric particles large regions in
parameter exist where lepton flavour violating decay modes have large
branching ratios despite the stringent constraints from the
non-observation of rare lepton decays.  We  discuss briefly some
consequences for discovery potential and the measurements of
edge-variables at LHC within the
SPS1a scenario. In the case of  R-parity violating
scenarios we focus on bilinear R-parity violation. We discuss
correlations between the decays of the lightest neutralino
and neutrino mixing angles as well as  the possibilities to
measure these correlations at LHC.
\end{abstract}

\section{Introduction}

The observed neutrino oscillations \cite{skam,sno,kland} are a clear
indication for non-vanishing neutrino masses and violation of
individual lepton numbers. 
Supersymmetry (SUSY) offers many possibilities to describe the observed neutrino 
data. The most popular one is certainly the usual seesaw mechanism
\cite{Minkowski:1977sc}, which 
introduces heavy right-handed neutrinos carrying a $\Delta L=2$ lepton
number violating Majorana mass. In the Minimal Supersymmetric Standard
Model (MSSM) a large $\nu_\mu$-$\nu_\tau$ mixing can lead to
a large $\tilde \nu_\mu$-$\tilde \nu_\tau$ mixing via renormalisation
group equations \cite{ref10}. An additional source of lepton flavour violation 
(LFV) arise in models where the MSSM with R-parity
conservation is embedded in a GUT theory 
\cite{Barbieri:1995tw}. This is a consequence of having
leptons and quarks in the same GUT multiplet. The quark flavour
mixing due to the CKM matrix leaves its traces also in the leptonic sector
\cite{Barbieri:1995tw,Ciuchini:2003rg}.

Therefore, one expects flavour violating effects for charged
leptons. Furthermore, in analogy to quarks, lepton flavour violation
may also be related to CP violation.
Lepton flavour violation (LFV) in the
charged lepton sector is, however, 
severely constrained by the stringent experimental
bounds on the branching ratios $BR(\mu \to e \gamma) < 1.2 \cdot 10^{-11}$,
$BR(\tau \to e \gamma) < 2.7 \cdot 10^{-6}$, $BR(\tau \to \mu \gamma) < 1.1
\cdot 10^{-6}$ and rare processes such as $\mu-e$ conversion \cite{pdg}.
Nevertheless,  clear LFV signals are expected in
slepton and sneutrino production and in the decays of neutralinos,
charginos, sleptons and sneutrinos at the LHC and at future lepton
colliders
\cite{ref12,Hinchliffe:2001np,ref8} despite 
these stringent constraints.  We will discuss such scenarios in 
the first half of this report.

Supersymmetry offers an interesting 
option to accommodate for the observed neutrino data
which is intrinsically supersymmetric: the breaking of R-parity. 
Adding bilinear terms to the MSSM superpotential is the simplest way
to realize this idea in practice. It has been shown that in this way
neutrino data can by successfully explained (see e.g.~\cite{NuMass}
and references therein). Moreover it has been demonstrated that
various decay properties of the lightest supersymmetric particle (LSP)
are correlated with neutrino properties, in particular with neutrino
mixing angles
\cite{NtrlDecay,NtrlOthers}. We will discuss such correlations taking
the lightest neutralino as LSP as well as the possibilities
to measure these correlations at the LHC.

\section{The R-parity conserving MSSM}

We will first discuss the case of conserved R-parity where total lepton
number is conserved but individual lepton is violated. In the absence
of right-handed neutrinos one can work without loss of generality in a basis
where the lepton Yukawa couplings are real and diagonal. In this basis
the complete information on LFV is encoded in the mass matrices of the
charged sleptons and of the sneutrinos:
\begin{equation}
  M^2_{\tilde l} = \left(
    \begin{array}{cc}
      M^2_{LL} &  M^{2\dagger}_{LR} \\
      M^2_{LR} &  M^2_{RR} \\
     \end{array} \right) \, , \,\,\,
  \label{eq:sleptonmass}
  M^2_{\tilde \nu,ij} =  M^2_{L,ij} \textstyle
  + \frac{1}{8} \left( g^2 + {g'}^2 \right) (v^2_d - v^2_u) \delta_{ij}
\end{equation}
where the entries in $M^2_{\tilde l}$ are $3 \times 3$ matrices that are given by
\begin{eqnarray}
  \label{eq:massLL}
  M^2_{LL,ij} &=& M^2_{L,ij} + \textstyle
\frac{1}{2} (v_d  Y_{i}^E)^2 \delta_{ij}
%  \nonumber \\ &&
  + \frac{1}{8}\left( {g'}^2 -  g^2 \right) (v^2_d - v^2_u) \delta_{ij}  \, ,\\
  \label{eq:sleptonmassLR}
  M^2_{LR,ij} &=&\textstyle 
       \frac{1 }{\sqrt{2}} \left( v_d A^*_{ij} - \mu v_u Y^E_{i} \delta_{ij}
          \right)  \, ,\\ 
  M^2_{RR,ij} &=& M^2_{E,ij} \textstyle 
    + \frac{1}{2}  (v_d  Y_{i}^E)^2 \delta_{ij}
      -  \frac{1}{4} {g'}^2  (v^2_d - v^2_u) \delta_{ij}  \, .%\\
\end{eqnarray}
 $M^2_{L}$ and $M^2_{E}$ are the soft SUSY breaking mass matrices for
left and right sleptons, respectively, and the $A_{ij}$ are the trilinear soft
SUSY breaking couplings of the sleptons and Higgs boson,
$\mu$ and the $Y^E_{l}$ are the usual $\mu$ parameter and 
the lepton Yukawa couplings such that $m_l = v_d Y^E_{l} / \sqrt{2}$. 
$v_u$ and $v_d$ are
the vacuum expectation values of the neutral Higgs fields (with 
$\tan\beta= v_u/v_d$).

\begin{figure}[t]
\setlength{\unitlength}{1mm}
\begin{picture}(150,57)
\put(0,-28){\mbox{\epsfig{figure=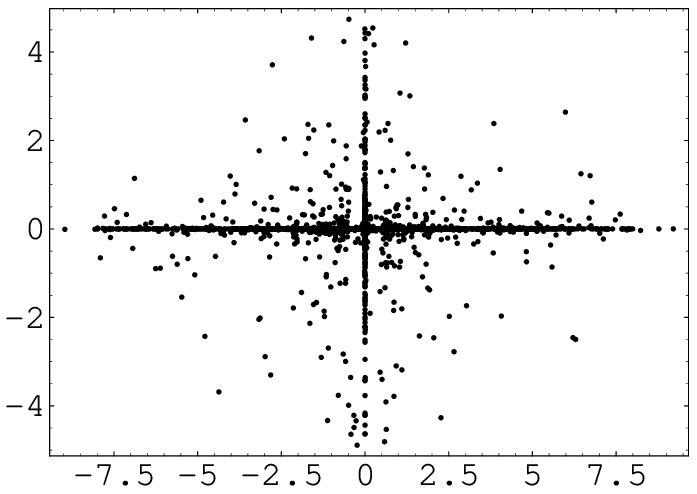,height=10.5cm,width=5.7cm}}}
\put(67,-28){\mbox{\epsfig{figure=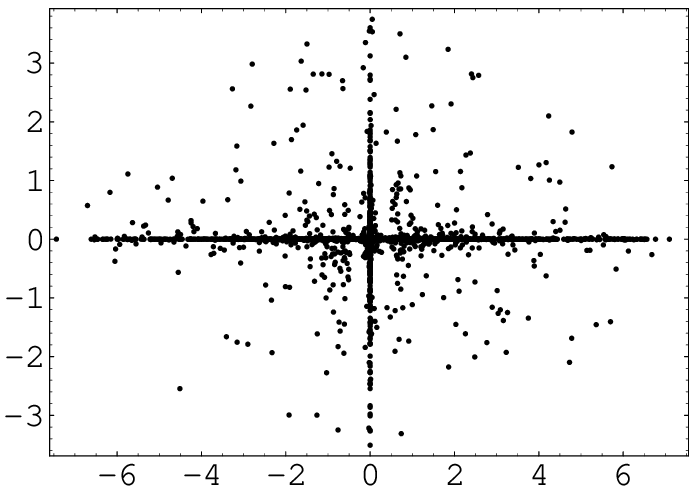,height=10.5cm,width=5.7cm}}}
\put(-2,53){\mbox{\bf a)}} 
\put(3,53){\mbox{$M^2_{L,13} \cdot 10^3$~GeV$^2$}}
\put(32,-4){\mbox{$M^2_{E,13} \cdot 10^3$~GeV$^2$}}
\put(65,53){\mbox{\bf b)}} 
\put(69,53){\mbox{$M^2_{L,23} \cdot 10^3$~GeV$^2$}}
\put(98,-4){\mbox{$M^2_{E,23} \cdot 10^3$~ GeV$^2$}}
\end{picture}
\caption{Ranges for parameters inducing lepton number violation.}
\label{fig:parameter}
\end{figure}

In the following we are interested in the effect of the off-diagonal
entries in the matrices  $M^2_{L}$, $M^2_{E}$ and $A_{ij}$. For this reason
we fix the diagonal entries of these matrices as well as the other
supersymmetric parameters by using the original high scale definition
of the Snowmass point SPS\#1a \cite{Allanach:2002nj}: 
$M_0=100$~GeV, $M_{1/2}=250$~GeV,
$A_0=-100$~GeV, $\tan\beta=10$ and $\mu>0$. At the electroweak scale
typical parameters are given as  
$M^2_{L,11} = 202.3^2$~GeV$^2$,
$M^2_{L,33} = 201.5^2$~GeV$^2$, $M^2_{E,11} = 138.7^2$~GeV$^2$, 
$M^2_{E,33} = 136.3^2$~GeV$^2$,
 $A_{11} = -7.567 \cdot 10^{-3}$~GeV, $A_{22} = -1.565$~GeV,
$A_{33} = -26.326$~GeV. To these parameters we add off-diagonal
elements such that the bounds from rare lepton decays are fulfilled.
We find values for  $|M^2_{E,ij}|$ up to $8 \cdot 10^3$~GeV$^2$, $|M^2_{L,ij}|$
up to $6 \cdot 10^3$~GeV$^2$ and $|A_{ij} v_d|$ up to 650~GeV$^2$ compatible 
with the 
constraints. In most cases, one of the mass squared parameters is at least
one order of magnitude larger than all the others. However, there is a
sizable part in parameters where at least two of the off-diagonal parameters
have the same order of magnitude as shown in Fig.~\ref{fig:parameter}.

\begin{figure}[t]
\setlength{\unitlength}{1mm}
\begin{picture}(150,59)
\put(-1,-3){\mbox{\epsfig{figure=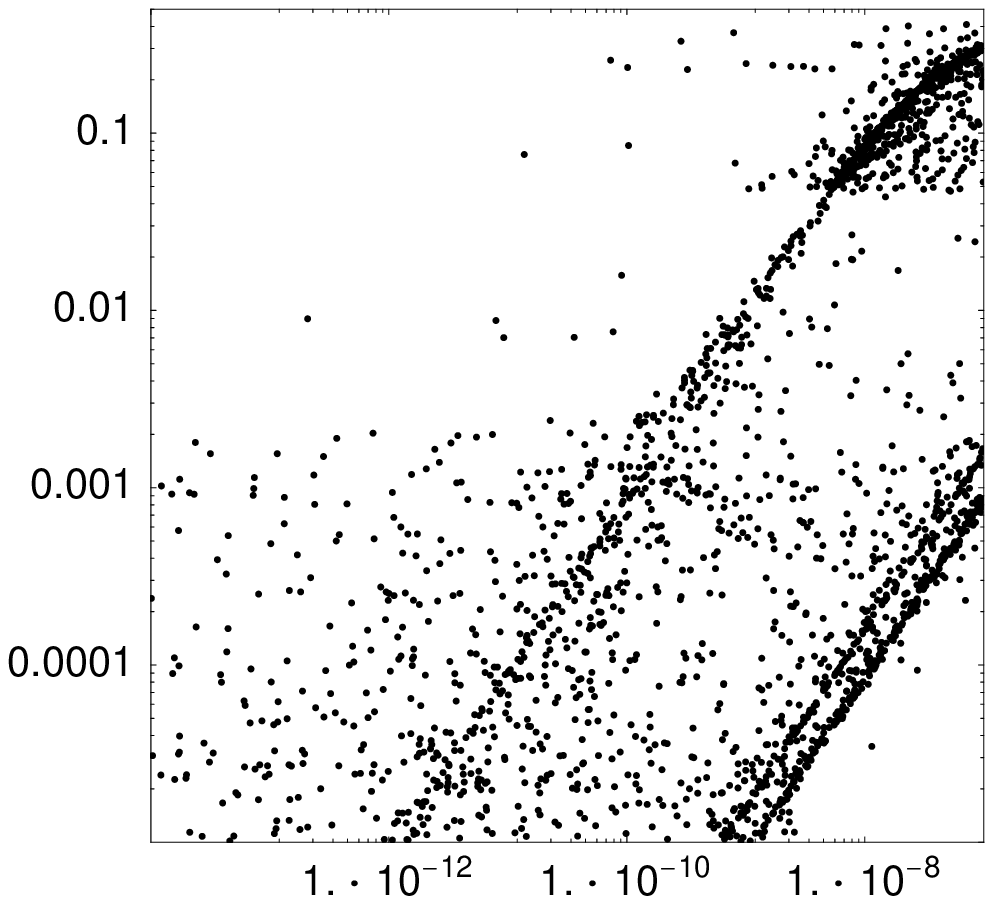,height=5.7cm,width=6.cm}}}
\put(65,-3){\mbox{\epsfig{figure=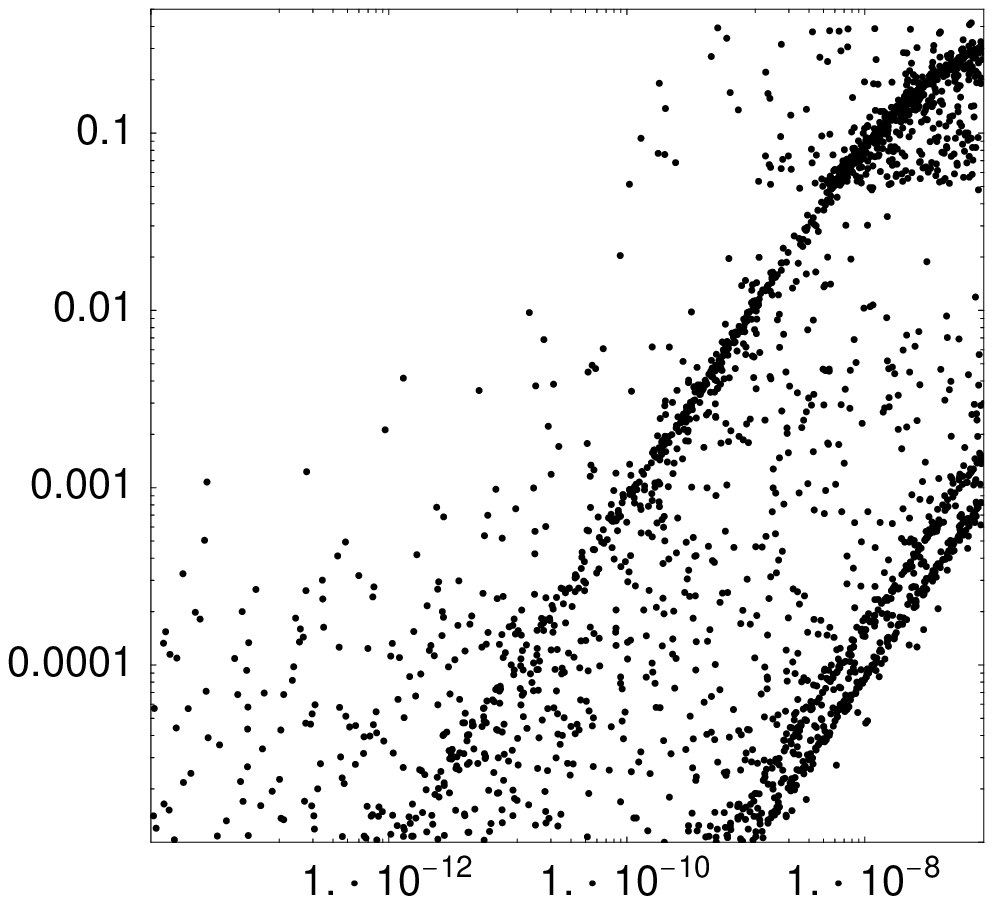,height=5.7cm,width=6.cm}}}
\put(-2,54){\mbox{\bf a)}} 
\put(2,54){\mbox{ 
           BR$({\tilde \chi}^0_2 \to{\tilde \chi}^0_1 \, e^\pm \, \tau^\mp)$}}
\put(39,-4){\mbox{  BR$(\tau \to e \gamma)$}}
\put(65,54){\mbox{\bf b)}} 
\put(69,54){\mbox{
         BR$({\tilde \chi}^0_2 \to{\tilde \chi}^0_1 \, \mu^\pm \, \tau^\mp)$}}
\put(105,-4){\mbox{ BR$(\tau \to \mu \gamma)$}}
\end{picture}
\caption{Branching ratios for lepton number violating decays of the second
     lightest neutralinos as a function of branching ratios of rare lepton
     decays taking SPS1a as starting point and adding lepton
     flavour violating off diagonal entries.}
\label{fig:chi2decays}
\end{figure}

These parameters induce lepton number violating couplings to
charginos and neutralinos  which in turn
lead lepton number violating decays such as 
$\tilde \chi^0_2 \to \tilde e_R \tau$, $\tilde e_R \to \mu \tilde \chi^0_1$ or
$\tilde e_R \to \tau \tilde \chi^0_1$. In Fig.~\ref{fig:chi2decays} we
show examples of lepton flavour violating decays modes of the second
lightest neutralino where we have summed over the intermediate slepton
states. As can be seen these branching ratios can go up to 40\%.
It has been shown in \cite{Hinchliffe:2001np} that lepton number violating
decay chains can be identified despite the considerable background processes
stemming from decays of 
supersymmetric particles. This implies that lepton number violation
in supersymmetric decays can be explored at the LHC.
In addition these decay modes are important for the LHC when one considers 
the so-called edge variables discussed in \cite{HinPai}. These variables
are extracted from decay chains such as 
$\tilde q_L \to q \tilde \chi^0_2 \to q l^\pm \tilde l_R^\mp
            \to q l^+ l^-  \tilde \chi^0_1$
and are very useful to measure not only mass differences of supersymmetric
particles but also the mass of the lightest supersymmetric particle. 
The effect of the LFV modes is two-fold: On the one hand the reduce
final states containing $e^+ e^-$ and $\mu^+ \mu^-$ pairs which might
deteriorate the accuracy of the mass measurements of selectrons and smuons.
On the other hand the could increase the accuracy of the stau mass measurements
as the final states do not not only contain $\tau^+ \tau^-$ pairs 
but also $e^\pm \tau^\mp$ and $\mu^\pm \tau^\mp$.

Finally we want to remark that in scenarios with large lepton number violation
the discovery reach of LHC could be enlarged taking SPS1a as example.
 In this scenario the
discovery of the lightest chargino is very difficult if not impossible
at LHC \cite{Giacomo} because the chargino decays with nearly 100\%
as follows: 
$\tilde \chi^+_1 \to \nu_\tau \tilde \tau_1^+
                \to \nu_\tau \tau^+ \tilde \chi^0_1$. In case of
sizable lepton number violation in the right slepton sector the lighter
stau can have large branching ratios into $e$ or $\mu$ final states. We have
found that in the above scanned parameter space the sum of the branching
ratios BR$(\tilde \tau_1 \to e \chi^0_1)$ + BR$(\tilde \tau_1 \to \mu \chi^0_1)$
can go up to 20\% even in case where the branching ratios for
rare $\tau$ decays are at a level of $10^{-10}$. This could enhance
considerably the discovery potential for the lightest chargino in this
scenario.

\section{Bilinear R-parity violation}

In supersymmetric theories Majorana mass terms for left-handed neutrinos
can be induced by introducing lepton-number and, thus, R-parity breaking
terms in the superpotential.
Adding bilinear terms to the MSSM superpotential is the simplest way
to realize this idea in practice explaining successfully
neutrino data (see e.g.~\cite{NuMass}
and references therein). In such a scenario an effective seesaw mechanism
takes place where the neutralinos play the role of the right-handed neutrinos.
This implies that
various decay properties of the lightest supersymmetric particle (LSP)
are correlated with neutrino properties, in particular with neutrino
mixing angles
\cite{NtrlDecay,NtrlOthers}. 

The Lagrangian of the model is obtained by adding bilinear
terms breaking lepton number to the MSSM  superpotential
and consistently the corresponding terms to the soft SUSY breaking potential:
\begin{eqnarray}
W_{\rm{BRpV}} = W_{\rm{MSSM}}  - \varepsilon_{ab}
\epsilon_i \widehat L_i^a\widehat H_u^b 
\hspace{5mm},\hspace{5mm}
V_{\rm{soft}} = V_{\rm{soft,MSSM}} -\varepsilon_{ab}
          B_i\epsilon_i\widetilde L_i^aH_u^b \, .
\end{eqnarray}
The latter induce vacuum expectation values $v_i$ for the sneutrinos which
are in turn responsible for mixing between standard model particles with
supersymmetric particles.
The mixing of neutrinos with neutralinos gives rise to one massive neutrino
at tree level. The other two neutrinos obtain masses due to loop
effects \cite{NuMass}. Assuming that the heaviest neutrino obtains its 
mass at tree level, the main features relevant for our current purpose are 
the following:
\begin{eqnarray}
 && \tan \theta_{\rm atm} = \left|\frac{\Lambda_2}{\Lambda_3}\right|
\hspace{4mm},\hspace{4mm} 
 \tan \theta_{\rm sol} \simeq
          \left|\frac{\tilde \epsilon_1}{ \tilde \epsilon_2}\right|
\hspace{4mm},\hspace{4mm} 
U_{e3}^2 \simeq \frac{\Lambda_1^2}{\Lambda_2^2 + \Lambda_3^2}
\label{eq:mixing} \\
&& \Lambda_i = \epsilon_i v_d + \mu v_i \, .
\hspace{4mm},\hspace{4mm} 
  \tilde \epsilon_i = V^{\nu,\rm tree}_{ij} \epsilon_j 
\end{eqnarray}
where $\theta_{\rm atm}$ is the atmospheric neutrino mixing angle, 
$\theta_{\odot}$ is the solar neutrino mixing angle and  
$V^{\nu,\rm tree}$ is the tree level neutrino mixing 
matrix \cite{NuMass}. 
% For a more thorough discussion see ref.~\cite{NuMass}. 

\begin{figure}[t]
\setlength{\unitlength}{1mm}
\begin{picture}(150,62)
\put(0,-3){\mbox{\epsfig{figure=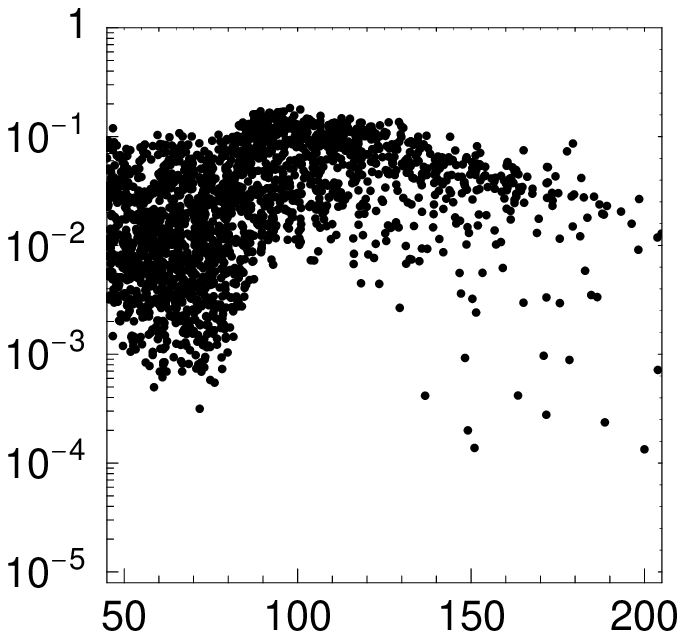,height=6.1cm,width=6.1cm}}}
\put(66,-3){\mbox{\epsfig{figure=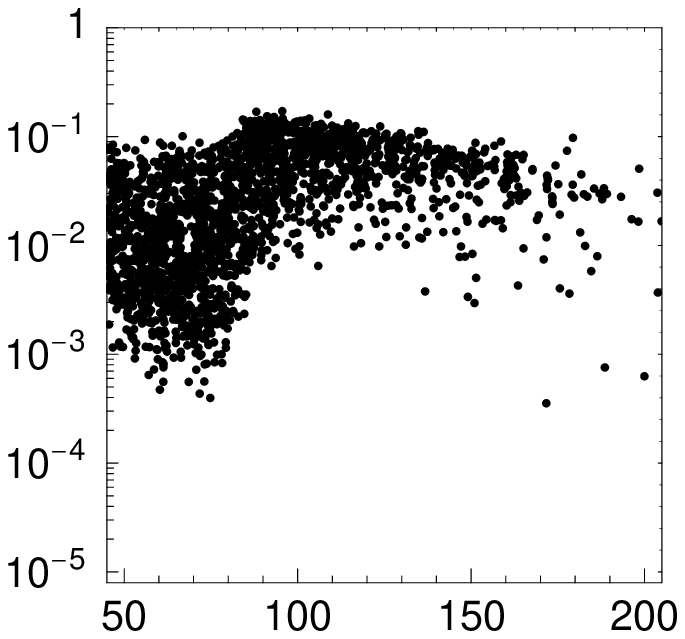,height=6.1cm,width=6.1cm}}}
\put(-2,58){\mbox{\bf a)}} 
\put(2,58){\mbox{ 
           BR$({\tilde \chi}^0_1 \to \mu^\mp q q')$}}
\put(44,-4){\mbox{$m_{\tilde \chi^0_1}$~[GeV]}}
\put(65,58){\mbox{\bf b)}} 
\put(69,58){\mbox{
         BR$({\tilde \chi}^0_1 \to \tau^\mp q q')$}}
\put(110,-4){\mbox{$m_{\tilde \chi^0_1}$~[GeV]}}
\end{picture}
\caption{Examples of branching ratios for semi-leptonic decays of the
lightest neutralino as a function of  $m_{\tilde \chi^0_1}$ scanning over
the SUSY parameter space.}
\label{fig:brchi1}
\end{figure}

In this model the neutrino spectrum is hierarchical and hence the
neutrino mass scales squared coincide with the the experimentally
measured neutrino mass squared 
differences. This implies that
the R-parity violating parameters are significantly
smaller than the R-parity conserving ones:
$|\epsilon_i| \ll |\mu|$ and $|v_i| \ll v_d$. This feature allows for
the possibility that all R-parity violating couplings can be expanded
in terms of the ratios\cite{NtrlDecay,NuMass}
\begin{eqnarray}
 \frac{\epsilon_i}{\mu}, \, \,
 \frac{\Lambda_i}{\sqrt{{\rm Det} (\tilde \chi^0)}} \hskip2mm \rm{or}\hskip2mm
 \frac{\Lambda_i}{{\rm Det}(\tilde \chi^+)} \, .
\end{eqnarray}
This implies that the neutrino mixing angles
 in Eq.~\ref{eq:mixing} can be expressed in ratios of couplings which
themselves are related to ratios of branching ratios.

\begin{figure}[t]
\setlength{\unitlength}{1mm}
\begin{picture}(150,62)
\put(0,-3){\mbox{\epsfig{
        figure=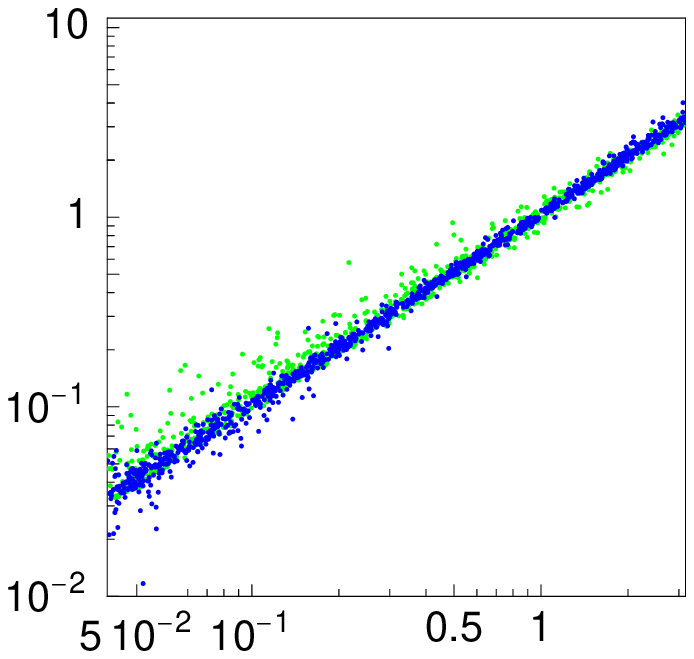,height=5.9cm,width=5.9cm}}}
\put(66,-3){\mbox{\epsfig{
          figure=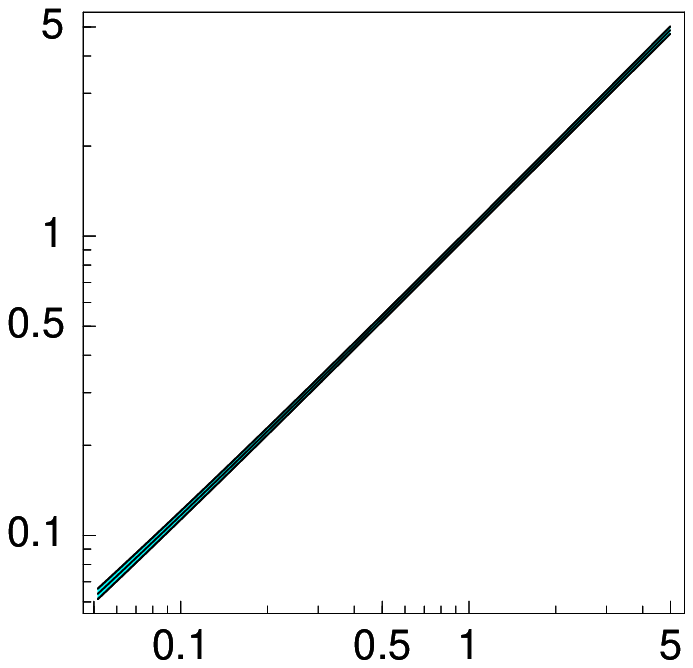,height=5.9cm,width=5.9cm}}}
\put(-2,58){\mbox{\bf a)}} 
\put(2,58){\mbox{ 
           BR$({\tilde \chi}^0_1 \to \mu^\mp q q')$ 
         / BR$({\tilde \chi}^0_1 \to \tau^\mp q q')$}}
\put(46,-4){\mbox{$\tan^2 \theta_{\rm atm}$}}
\put(64,58){\mbox{\bf b)}} 
\put(67,58){\mbox{
         BR$({\tilde \chi}^0_1 \to \mu^\mp q q')$ 
         / BR$({\tilde \chi}^0_1 \to \tau^\mp q q')$}}
\put(112,-4){\mbox{$\tan^2 \theta_{\rm atm}$}}
\end{picture}
\caption{Ratios of branching ratios of semi-leptonic neutralino decays as
a function of the $\tan^2 \theta_{\rm atm}$. In {\bf a)} the predictions
are shown without assuming any knowledge on the supersymmetric parameters
whereas in {\bf b)} it has assumed that the parameters are know with a precision
of 10\%.}
\label{fig:corr}.
\end{figure}

In the following we concentrate on the case that the lightest neutralino
is the LSP. 
We have performed a scan over the MSSM parameter space and added
 the R-parity violating parameters  such, that
$\Delta^2_{\rm atm}$ and  $\Delta^2_{\rm sol}$ and
at least two of the three neutrino mixing angles are in the experimentally
allowed range. In Fig.~\ref{fig:brchi1} the branching ratios 
 BR$({\tilde \chi}^0_1 \to \mu^\mp q q')$ and
BR$({\tilde \chi}^0_1 \to \tau^\mp q q')$ are shown as a function of
$m_{\tilde \chi^0_1}$. One sees that these branching ratios are in general
in the range of a few per-mile up to about 20\%. The importance of these
decay modes is that they are correlated with the atmospheric neutrino mixing
angle as shown in Fig.~\ref{fig:corr}. In Fig.~\ref{fig:corr}a we 
show the predictions of this model for the ratio
BR$({\tilde \chi}^0_1 \to \mu^\mp q q')$/BR$({\tilde \chi}^0_1
 \to \tau^\mp q q')$ scanning of the SUSY parameter space yielding a
clear correlation with $\tan^2 \theta_{\rm atm}$. The band collapses
to a line if the SUSY parameters are known as shown in
 Fig.~\ref{fig:corr}b. Here we have assumed that the SUSY parameters
are known with a precision of 10\% and we have taken into account the statistical
error on these branching ratios 
assuming $10^5$ identified neutralinos. In particular the masses
and mixing angles of neutralinos, sbottoms and staus are important in this
context \cite{NtrlDecay}.

In ref.~\cite{Porod:2004rb} a Monte Carlo study has been performed where
the ratio BR$({\tilde \chi}^0_1 \to \mu^\mp q q')$/BR$({\tilde \chi}^0_1
 \to \tau^\mp q q')$ has been investigated
within the SPS1a scenario adding R-parity violating parameters. Assuming
an integrated luminosity of 100 fb$^{-1}$ 
it has been shown that this ratio can indeed
be measured with a precision of about three per-cent. This clearly shows that
LHC is capable to test at least part of these correlations.
The main ingredients
for this statement are: (i) The considered semi-leptonic decay modes
have a branching ratio of a few per-cent. (ii) The neutralino has
a visible decay length. The latter is in particular useful to suppress
background stemming from SM and SUSY processes.

\section{Summary}

We have discussed briefly the possibilities to study lepton number violating
process at LHC in the context of supersymmetric theories. We have seen
that LHC can explore lepton number violation in the decays
of supersymmetric particles 
independent whether R-parity is conserved or violated. In the case of
R-parity conservation we have commented on the effects of
lepton flavour violating decay modes on edge variables and the
discovery potential of LHC. In case of R-parity violation we have
pointed out that it should be possible to measure at the LHC
correlations between neutrino mixing angles and ratios of LSP
branching ratios.

\bigskip

{\small This work was supported by a Spanish MCyT Ramon y Cajal contract,
in part by the Spanish grant BFM2002-00345 and
 by the European Commission RTN network HPRN-CT-2000-00148.}

%\lastevenpage
\end{document}